# Superconductivity in non-centrosymmetric rhombohedral NbSe$_2$

Zhengxian Li[1†], Xiaoyu Shen[1†], Kai Liu[1], Yating Sha[1], Tianyang Wang[1], Feng Liu[1], Qingchen Duan[1], Kenji Watanabe[2], Takashi Taniguchi[3], Peng Chen[1], Shiyong Wang[1], Ruidan Zhong[1], Dong Qian[1], Shengwei Jiang[1], Yufan Li[4], Noah F. Q. Yuan[1], Guorui Chen[1*]

[1]*State Key Laboratory of Micro-nano Engineering Science, Key Laboratory of Artificial Structures and Quantum Control (Ministry of Education), Tsung-Dao Lee Institute and School of Physics and Astronomy, Shanghai Jiao Tong University, Shanghai, China*

[2]*Research Center for Electronic and Optical Materials, National Institute for Materials Science, 1-1 Namiki, Tsukuba, Japan*

[3]*Research Center for Materials Nanoarchitectonics, National Institute for Materials Science, 1-1 Namiki, Tsukuba, Japan*

[4]*Department of Physics, The Chinese University of Hong Kong, Hong Kong, China*

†These authors contributed equally to this work.
*Correspondence to: chenguorui@sjtu.edu.cn

**Abstract:** Crystal stacking offers a powerful yet underexplored route to engineer symmetry in layered superconductors. Here we report superconductivity in rhombohedral-stacked NbSe$_2$ (3R-NbSe$_2$), a non-centrosymmetric polytype in which global inversion symmetry is removed by stacking alone. Using comprehensive structural, transport, magnetic, and thermodynamic measurements, we establish superconductivity as a bulk property of the 3R phase and find that the in-plane upper critical field exceeds the Pauli paramagnetic limit, indicating the persistence of strong Ising-type spin–orbit coupling. Unlike the thickness-dependent superconductivity in centrosymmetric 2H-NbSe$_2$, the superconducting transition temperature in 3R-NbSe$_2$ shows little dependence on layer number but exhibits an unusually strong sensitivity to disorder. We further observe strongly enhanced nonlinear optical and electrical responses near the superconducting transition, consistent with stacking-induced inversion-symmetry breaking. Our results identify 3R-NbSe$_2$ as a single-phase platform in which stacking engineering reshapes superconductivity and enables nonlinear transport phenomena in layered materials.

**Introduction**

The symmetry of a crystal lattice plays a central role in shaping collective quantum states of electrons, particularly in superconductors where spin, orbital, and pairing degrees of freedom are strongly intertwined[1,2]. Layered transition-metal dichalcogenides (TMDs) provide an ideal platform to examine how crystal symmetry and spin–orbit coupling (SOC) shape superconductivity, owing to their low dimensionality, strong intrinsic SOC, and multiple stacking polytypes[3-8]. Among them, NbSe$_2$ has emerged as a prototypical system in which superconductivity can be systematically tuned by thickness, magnetic field, and crystal structure[9-12]. In the widely studied 2H polytype, the bulk crystal is centrosymmetric despite the absence of inversion symmetry in each monolayer, giving rise to strong out-of-plane Ising-type SOC (ISOC) that suppresses Zeeman pair breaking and enables in-plane upper critical fields beyond the Pauli limit[9,13,14].

The rhombohedral (3R) polytype of NbSe$_2$ provides a minimal platform for removing global inversion symmetry by stacking alone. As illustrated in Fig. 1a, the difference between the 2H and 3R structures lies in the relative registry of adjacent NbSe$_2$ layers: in the 2H polytype, neighboring layers are rotated by 180°, restoring inversion symmetry in the bulk, whereas in the 3R polytype the layers retain the same in-plane orientation and follow an ABC stacking sequence. This stacking geometry removes the inversion center throughout the crystal and yields a non-centrosymmetric bulk structure. As a result, antisymmetric SOC (ASOC), absent in the centrosymmetric 2H polytype, becomes allowed[1]. This symmetry breaking is achieved without changing chemical composition or introducing interfaces, making 3R-NbSe$_2$ an intrinsic platform to examine how stacking-induced inversion-symmetry breaking reshapes superconductivity.

Here we investigate superconductivity in non-centrosymmetric rhombohedral NbSe$_2$ as a stacking-engineered counterpart of the well-established Ising superconductor 2H-NbSe$_2$. We synthesize 3R-NbSe$_2$ single crystals and demonstrate that superconductivity is a bulk property of the rhombohedral phase, with an in-plane upper critical field exceeding the Pauli paramagnetic limit, consistent with the persistence of strong ISOC. By systematically comparing samples with different disorder levels, we find that the superconducting transition temperature is largely independent of thickness but exhibits a pronounced correlation with disorder, as quantified by the residual resistivity ratio (RRR). In addition, we observe strongly enhanced nonlinear responses in the superconducting state, including optical second-harmonic generation and electrical second-harmonic transport, which are absent or much weaker in the centrosymmetric 2H phase. Together, these results establish 3R-NbSe$_2$ as a minimal platform to explore how stacking-induced inversion-symmetry breaking reshapes Ising superconductivity and enables nonlinear responses in layered superconductors.

**Crystal growth and structural characterization**

Rhombohedral NbSe$_2$ single crystals were grown by the chemical vapor transport method using iodine as the transport agent (see Methods). The as-grown crystals exhibit a flat, plate-like morphology typical of layered transition-metal dichalcogenides. Structural characterization was performed using complementary chemical, reciprocal-space, and real-space probes to determine both composition and crystal structure.

Energy-dispersive X-ray spectroscopy (EDS) measured over multiple spots reveals a slightly Nb-rich composition in the rhombohedral samples (see Fig. S1 and Table S1), in contrast to the near-stoichiometric composition typically observed in 2H-NbSe$_2$ (Fig. S2 and Table S2). Such Nb excess has been reported to favor the stabilization of the metastable rhombohedral stacking during crystal growth.[15]

To determine the crystallographic phase and exclude possible polytype mixing, we performed X-ray diffraction (XRD) measurements on the *ab* surface of the as-grown crystals. The diffraction pattern (Fig. 1b) shows a series of sharp (00*l*) reflections, indicating high crystallinity and a well-defined orientation with the c axis perpendicular to the surface. The positions of these peaks are consistent with the expected interlayer spacing of 3R-NbSe$_2$. Single-crystal XRD measurements further confirm that the structure belongs to the non-centrosymmetric *R*3*m* space group, with lattice parameters $a = b = 3.472$ Å and $c = 18.86$ Å (see Table S3). These results unambiguously establish the rhombohedral stacking symmetry of the crystals.

Local stacking order was further examined using high-angle annular dark-field scanning transmission electron microscopy. Atomic-resolution images directly reveal an ABC stacking sequence along the *a* axis (Fig. 1d), characteristic of the rhombohedral structure. No stacking faults or rotational domains are observed within the imaged region, confirming that the crystals are single-phase and exhibit uniform 3R stacking at the nanoscale.

**Bulk superconductivity in rhombohedral NbSe$_2$**

Having established the rhombohedral crystal structure, we next demonstrate that superconductivity in 3R-NbSe$_2$ is a bulk property of the crystal. Figure 2a shows the temperature dependence of the electrical resistance *R* measured on a representative bulk crystal. In the normal state, the resistance exhibits an approximately linear temperature dependence over a wide temperature range, which is more pronounced than that typically reported for the centrosymmetric 2H polytype. Upon further cooling, a clear superconducting transition is observed, followed by the emergence of a zero-resistance state at lower temperatures. The superconducting transition temperature $T_c$ is defined as the temperature at which the resistance drops to 50% of its normal-state value, and reaches up to 6.5 K among the measured bulk crystals. The RRR (defined by the ratio of resistances at 300K and 10K in this study) of 3R phase is approximately 5, significantly lower than that typically reported for 2H-NbSe$_2$ (10 to 30)[14,16-22], indicating a higher level of disorder in rhombohedral crystals. This enhanced disorder is attributed to the Nb-rich growth conditions required to stabilize the 3R stacking, which introduce point defects and increase impurity scattering.

Bulk superconductivity is further confirmed by magnetic susceptibility measurements. Figure 2b shows the temperature dependence of dc magnetization measured with zero-field-cooling (ZFC) and field-cooling (FC) mode under an applied magnetic field $H$ = 20 Oe along the $c$ axis. A pronounced diamagnetic response develops below $T_c$, providing direct evidence for magnetic flux expulsion. The onset temperature of the diamagnetic signal coincides with the transport measurements, indicating a homogeneous superconducting phase throughout the crystal.

Thermodynamic evidence for bulk superconductivity is provided by specific-heat measurements. As shown in Fig. 2c, a sharp anomaly in $C_p/T$ is observed at approximately 6 K under zero magnetic field, confirming a bulk superconducting transition.

We further examine the response of the superconductivity to an in-plane magnetic field. Figure 2d shows the temperature dependence of the upper critical field $H_{c2}^{\parallel}$ for a bulk 3R-NbSe$_2$ crystal. Notably, $H_{c2}^{\parallel}$ exceeds the Pauli paramagnetic limit, $H_P \approx 1.84T_{c0}$ ($T_{c0}$ is $T_c$ at zero magnetic field)[23,24], indicating that strong ISOC remains effective in suppressing Zeeman pair breaking in the rhombohedral phase. This behavior is expected because the ISOC in NbSe$_2$ originates primarily from the local crystal field and strong atomic SOC of the Nb $d$ orbitals within each NbSe$_2$ layer, and is therefore largely insensitive to the interlayer stacking sequence[2,25,26]. The persistence of Pauli-limit violation in bulk 3R-NbSe$_2$ thus establishes an important baseline for comparison with the centrosymmetric 2H polytype and sets the stage for examining how global inversion-symmetry breaking modifies superconductivity beyond the Ising paradigm.

**$T_c$-RRR correlation**

For 2H-NbSe$_2$, $T_c$ decreases with reducing layer thickness, reflecting the evolution of dimensionality and ISOC from bulk to the monolayer limit[14]. It is therefore natural to ask whether a similar thickness dependence persists in the rhombohedral phase, where global inversion symmetry is absent at all thicknesses. As shown below, this expectation breaks down in 3R-NbSe$_2$.

Figure 3a shows the transport measurements of multiple 3R-NbSe$_2$ devices with thicknesses ranging from bulk down to bilayer. When $T_c$ is summarized as a function of layer number (Fig. 3b), no systematic thickness dependence is observed. This absence of a clear thickness trend indicates that dimensional confinement alone does not govern the superconducting transition in the rhombohedral phase, and suggests that an alternative parameter plays a more dominant role.

A qualitatively different picture emerges when disorder is taken into account. As shown in Fig. 3b, the RRR is plotted for different devices, where the $T_c$ and RRR is correlated for all thicknesses. By summarize $T_c$ and corresponding RRR in Fig. 3c, $T_c$ exhibits a pronounced, nearly linear correlation with RRR accross all measured 3R-NbSe$_2$ samples, independent of thickness. This behavior contrasts sharply with that reported for 2H-NbSe2 (blue symbols), where superconductivity is comparatively insensitive to RRR over a broad range of values[14,16-22]. The collapse of data from samples with different thicknesses onto a single $T_c$–RRR trend highlights disorder, rather than dimensionality, as the dominant tuning parameter for the superconductivity in the rhombohedral phase.

The strong disorder sensitivity observed in 3R-NbSe$_2$ points to a qualitative change in the superconducting state induced by stacking-driven inversion-symmetry breaking. In non-centrosymmetric superconductors, ASOC can admix pairing channels of different parity, rendering the superconducting state more susceptible to impurity scattering[1,2,25]. While a detailed microscopic mechanism lies beyond the scope of this work, our results establish disorder as an efficient and intrinsic control parameter for superconductivity in 3R-NbSe$_2$, highlighting a fundamental distinction from the conventional Ising superconductivity realized in the centrosymmetric 2H polytype.

**Enhanced nonlinear responses in 3R-NbSe$_2$**

The stacking-induced inversion-symmetry breaking in 3R-NbSe$_2$ naturally motivates a search for nonlinear responses that are forbidden or strongly suppressed in the centrosymmetric 2H polytype. Figure 4 summarizes the nonlinear optical and electrical responses measured in 3R-NbSe$_2$ and directly compares them with those of 2H-NbSe$_2$.

We first examine the optical second-harmonic generation (SHG). Figure 4a shows the measured SHG signals at room temperature for bulk 3R- and 2H-NbSe$_2$, where 3R-NbSe$_2$ exhibits a significantly enhanced SHG signal - more than two orders of magnitude larger than that of 2H-NbSe$_2$ - with a characteristic six-fold symmetry. This pronounced enhancement is consistent with the absence of global inversion symmetry in the rhombohedral stacking and provides an independent confirmation of the non-centrosymmetric crystal structure established in Fig. 1. In contrast, the SHG signal in bulk 2H-NbSe$_2$ remains weak, reflecting the restoration of inversion symmetry.

More strikingly, we observe pronounced nonlinear electrical responses in the superconducting state of 3R-NbSe$_2$. Figure 4b shows the temperature dependence of the second-harmonic voltage $V_{2\omega}$ measured under an ac current excitation in a thin-layer 3R-NbSe$_2$ device. The second-harmonic signal emerges only within the superconducting transition window and rapidly diminishes both above and well below $T_c$. This behavior indicates that the nonlinear electrical response is intrinsically tied to the superconducting state, rather than originating from normal-state transport or extrinsic effects such as contacts or Joule heating[27-33]. By comparison, the corresponding signal in 2H-NbSe$_2$ is more than two orders of magnitude smaller under comparable measurement conditions (Fig. 4c), underscoring the role of stacking-induced inversion-symmetry breaking.

Phenomenologically, we can construct the Ginzburg-Landau free energy for a uniform sample of superconductor in three dimensions

$$F = \int \alpha(\boldsymbol{q})|\psi_q|^2 d^3\boldsymbol{q}$$

with order parameter $\psi_q$, Cooper pair momentum $\boldsymbol{q}$ and coefficient $\alpha(\boldsymbol{q})$. Then within the superconducting transition window, one can find the electric current response under an electric field[34]

$$J_a = \sigma_{ab}E_b + \chi_{abc}E_b E_c$$

$$\sigma_{ab} = \frac{e^2 k_B T \gamma}{\hbar} \int \frac{\partial_{ab}\alpha}{\alpha^2} d^3\boldsymbol{q}$$

$$\chi_{abc} = \frac{2}{3}\frac{e^3 k_B T \gamma^2}{\hbar} \int \frac{\partial_{abc}\alpha}{\alpha^3} d^3\boldsymbol{q}$$

where $\partial_{ab} \equiv \partial^2/\partial q_a \partial q_b$, $\partial_{abc} \equiv \partial^3/\partial q_a \partial q_b \partial q_c$ and $\gamma$ is the damping parameter of Cooper pairs in the superconducting transition window.

When the inversion symmetry is preserved, $\alpha(-\boldsymbol{q}) = \alpha(\boldsymbol{q})$, $\partial_{abc}\alpha(-\boldsymbol{q}) = -\partial_{abc}\alpha(\boldsymbol{q})$, and $\chi_{abc} \equiv 0$ as the integrand is odd in momentum. When the inversion symmetry is broken, $\chi_{abc}$ can become nonzero. In this case, under a uniform current $I$, the longitudinal voltage drop reads

$$V = \frac{L}{\sigma S}I - \frac{\chi L}{\sigma^3 S^2}I^2$$

with length $L$ along current direction (denoted as $x$-axis) and cross-sectional area $S$ perpendicular to current direction, and $\sigma = \sigma_{xx}, \chi = \chi_{xxx}$.

In addition, the $V_{2\omega}$ can take either positive or negative values depending on the sample and measurement configuration. Such sign reversals are naturally expected for nonlinear responses that are sensitive to microscopic details, including current distribution, disorder landscape, and the relative orientation between current flow and crystal axes, rather than being fixed by symmetry alone. The confinement of strong nonlinear transport to the vicinity of the superconducting transition suggests an enhanced susceptibility of the superconducting state to inversion-symmetry breaking in 3R-NbSe$_2$. While a detailed microscopic mechanism remains to be clarified, these results demonstrate that stacking-engineered inversion-symmetry breaking manifests not only in static superconducting properties but also in pronounced nonlinear transport responses.

**Summary and outlook**

In summary, we have established rhombohedral NbSe$_2$ as a bulk superconductor in which crystal stacking alone removes global inversion symmetry while preserving strong ISOC. By disentangling the roles of thickness, disorder, and symmetry, our results show that superconductivity in the 3R phase is governed by a distinct interplay between stacking-induced inversion-symmetry breaking and disorder, and is accompanied by strongly enhanced nonlinear responses near the superconducting transition.

These findings position 3R-NbSe$_2$ as a single-phase platform for exploring superconductivity beyond the centrosymmetric Ising paradigm. Future studies combining phase-sensitive probes, spectroscopic measurements, and controlled disorder tuning may further clarify how inversion-symmetry breaking influences pairing, fluctuations, and nonlinear transport in layered superconductors. More broadly, our work highlights stacking engineering as an effective route to access symmetry-enabled superconducting responses without altering chemical composition or introducing interfacial complexity.

**Note added.** During the preparation of this manuscript, we became aware of a related study reporting superconductivity in sliding 3R-NbSe$_2$[35].

**Acknowledgements**

We thank Quansheng Wu and Xu Yan for valuable discussions, and Na Yu for assistance of single crystal XRD measurement. G.C. acknowledges support from NSF of China (grant no. 12350005), National Key R&D Program of China (grant no. 2021YFA1400100), and Shanghai Science and Technology Innovation Action Plan (grant no. 24LZ1401100). Z.L. acknowledges support from the Postdoctoral Fellowship Program of CPSF (grant no. GZC20241021). S.W. acknowledges NSF of China (grant no. 92577203). S.J. acknowledges National Key R&D Program of China (grant no. 2021YFA1401400), NSF of China (grant no. 12550403) and Yangyang Development Fund. R.Z. acknowledges National Key R&D of China (grant nos. 2022YFA1402702 and 2021YFA1401600), NSF of China (grant nos. 12334008 and 12374148). X.S. is supported by T.D. Lee scholarship.

**Methods**

**Crystal growth**

Single crystals of 2H- and 3R-NbSe$_2$ were grown by the chemical vapor transport (CVT) method using iodine as the transport agent. For 2H-NbSe$_2$, high-purity niobium powder (99.95%) and selenium granules (99.9%) with a slight excess of Se were sealed in an evacuated quartz tube together with iodine (<5 mg cm$^{-3}$). Approximately 1 g of the mixture was placed in a two-zone tubular furnace, with the source and growth zones maintained at 825 °C and 795 °C, respectively, for 250 h. High-quality plate-like single crystals were obtained at the cold end after slow cooling to room temperature.

For 3R-NbSe$_2$, niobium and selenium were mixed in a nominal molar ratio of 1:2 with a slight excess of Nb, which is required to stabilize the metastable rhombohedral stacking. The mixture was sealed in an evacuated quartz tube and placed in a two-zone furnace with a temperature gradient of 1050 °C (source) to 1020 °C (growth) for 250 h. After growth, the tube was rapidly quenched in iced water to preserve the rhombohedral phase. The as-grown crystals exhibit flat, plate-like morphology.

**Structure Characterization**

XRD measurements were performed to determine the crystallographic phase and structural quality. Single-crystal XRD data were collected at 293 K on a Bruker D8 diffractometer using Mo K$\alpha$ radiation ($\lambda$ = 0.71073 Å). Structure refinement confirms that the diffraction patterns are fully consistent with the rhombohedral space group $R3m$, with lattice parameters $a = b = 3.472$ Å and $c = 18.86$ Å ($\alpha = \beta = 90°$, $\gamma = 120°$). Detailed crystallographic information and refinement results are summarized in Table S3.

Powder XRD patterns measured on the *ab* surface were obtained using a Bruker D8 Advance diffractometer with Cu K$\alpha_1$ radiation ($\lambda$ = 1.54184 Å), showing only (00$l$) reflections characteristic of highly oriented layered crystals.

HAADF-STEM measurements were carried out on a Thermo Fisher Spectra 300 microscope. Atomic-resolution images reveal an ABC stacking sequence along the *c* axis, characteristic of the rhombohedral structure. No stacking faults or rotational domains were observed within the imaged regions, indicating uniform 3R stacking at the nanoscale and excluding polytype mixing.

Chemical composition was examined by EDS using a JEOL JSM-7800F scanning electron microscope. More than six randomly selected regions were measured on each crystal. The averaged composition yields Nb:Se ≈ 1.128:2 (Table S1), indicating a Nb-rich stoichiometry compared with typical 2H-NbSe$_2$ (Nb:Se ≈ 1.04:2). Such Nb excess is consistent with previous reports on stabilizing rhombohedral stacking during crystal growth.

**Thin flake and device fabrication**

3R-NbSe$_2$ and hBN flakes were mechanically exfoliated from bulk crystals onto SiO$_2$ (285 nm)/Si substrates using adhesive tape. Thin flakes were first identified under an optical microscope, and their layer numbers were determined by optical reflectance contrast calibrated with atomic force microscopy on selected samples. A dry-transfer technique based on a polymer stamp was used to assemble the heterostructures. Specifically, an hBN flake was first picked up by a polycarbonate (PC)/polydimethylsiloxane (PDMS) stamp and subsequently used to pick up the target 3R-NbSe$_2$ flake, forming an hBN/3R-NbSe$_2$ stack. The assembled heterostructure was then released onto pre-patterned Cr/Pd (3 nm/10 nm) electrodes on a SiO$_2$/Si substrate by heating the stage to soften the polymer layer, followed by removal of the polymer in chloroform. The electrodes were fabricated in advance by standard electron-beam lithography and lift-off processes.

**Specific heat measurements**

Specific heat measurements were performed on bulk 3R-NbSe$_2$ crystals using a relaxation method. The electronic specific heat coefficient $\gamma$ was extracted by fitting the normal-state data at 5 T to $C_p/T = \gamma + \beta T^2$, where $\beta T^3$ and $\gamma T$ are the phonon contribution and the electronic specific heat coefficient to the specific heat, respectively[36]. From the fit, we obtain $\gamma$ = 12.38 mJ mol$^{-1}$ K$^{-2}$. An equal-area construction applied to the superconducting anomaly yields $T_c$ = 5.95 K and a normalized specific heat jump $\Delta C/(\gamma T_c)$ = 1.52, slightly larger than the weak-coupling BCS value of 1.43[37]. The corresponding superconducting volume fraction is estimated to be close to 100%, confirming bulk superconductivity.

**Transport measurement**

Transport measurements were performed in Oxford variable temperature insert system and Physike system, from 0.3 K to 300 K, equipped with a helium-3 probe. Standard low-frequency lock-in techniques were used to measure the sample resistances with an ac excitation current of 100 nA-1mA at ~17.777 Hz.

Second-harmonic electrical signals were measured using a lock-in amplifier (Stanford Research Systems SR860). Under an ac excitation current $I^\omega = I_0 sin(\omega t)$, the second-order voltage contains both dc and second-harmonic components. We focus on the second-harmonic voltage $V_{2\omega}$, which enables phase-sensitive detection with enhanced sensitivity and minimizes contributions from linear transport.

**Second-harmonic generation (SHG) measurement.**

Angle-resolved second-harmonic generation (SHG) measurements were performed in a reflection geometry. The excitation source was an 800 nm femtosecond laser (Chameleon Ultra II, Coherent), which was linearly polarized and intensity-modulated with an optical chopper (MC2000B, Thorlabs). An achromatic half-wave plate mounted on a high-precision rotation stage (PRM1Z8, Thorlabs) was used to control the excitation polarization relative to the sample orientation. The reflected SHG signal passed back through the same half-wave plate and then through an analyzer set perpendicular to the excitation polarization. A short-pass filter in the detection path was used to suppress the fundamental wavelength. The SHG was detected with a photomultiplier tube using a standard lock-in technique (SR830, Stanford Research Systems).

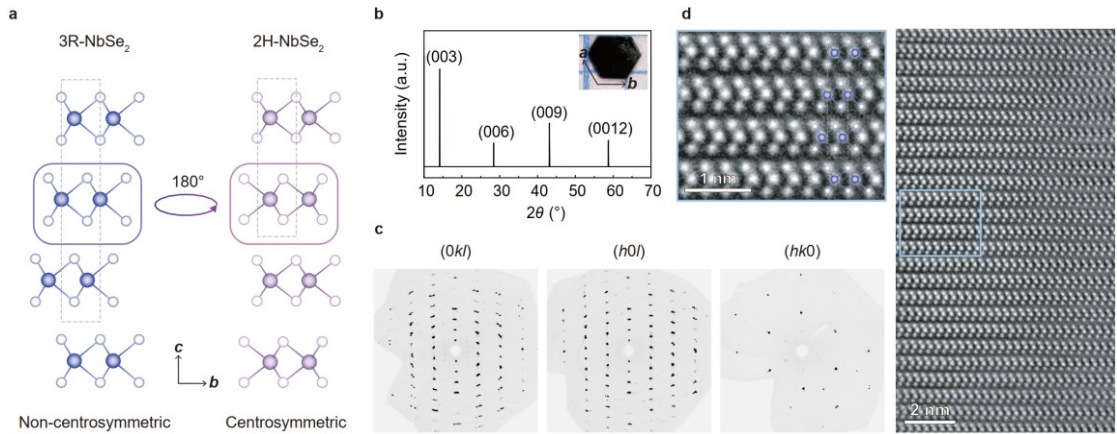

**Figure 1. Crystal structure of non-centrosymmetric rhombohedral NbSe$_2$.** **(a)** Schematic crystal structures of 3R-NbSe$_2$ (left) and 2H-NbSe$_2$ viewed along the *a* axis. In the 2H polytype, adjacent layers are rotated by 180°, restoring inversion symmetry in the bulk. In contrast, in the 3R polytype the layers retain the same in-plane orientation and stack in an ABC sequence, resulting in a non-centrosymmetric bulk structure. Solid spheres represent Nb atoms, and open circles represent Se atoms. **(b)** Room-temperature X-ray diffraction (XRD) pattern measured on the *ab* surface of a 3R-NbSe$_2$ single crystal, showing sharp (00*l*) reflections characteristic of a highly oriented layered structure. Inset: optical image of a representative 3R-NbSe$_2$ single crystal synthesized in this work. **(c)** Single-crystal XRD reciprocal-space maps of 3R-NbSe$_2$ measured along the (0*kl*), (*h*0*l*), and (*hk*0) directions, confirming the high crystalline quality and rhombohedral symmetry. **(d)** High-angle annular dark-field scanning transmission electron microscopy (HAADF-STEM) image of a 3R-NbSe$_2$ single crystal. The projected atomic columns along the *a* axis (*bc* plane) reveal the characteristic ABC stacking sequence of the rhombohedral structure. The region highlighted by the blue rectangle is shown at higher magnification.

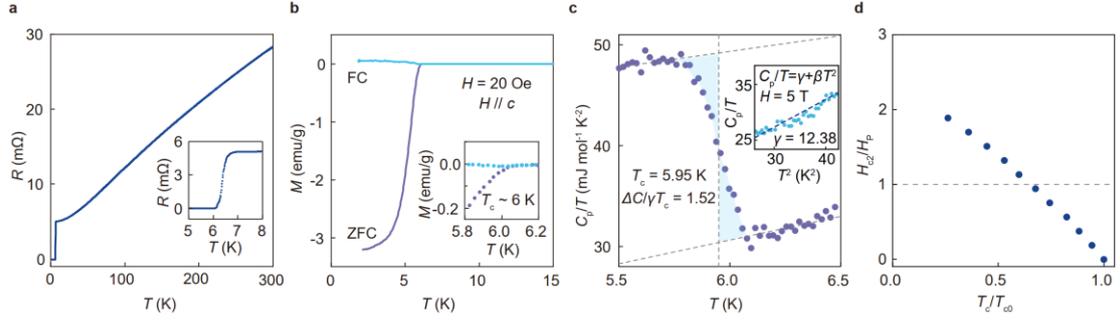

**Figure 2. Bulk superconductivity in rhombohedral NbSe$_2$.** (a) Temperature-dependent electrical resistance of a bulk 3R-NbSe$_2$ crystal. Inset: enlarged view of the low-temperature region highlighting the superconducting transition. (b) DC magnetization of 3R-NbSe$_2$ measured in zero-field-cooling (ZFC) and field-cooling (FC) modes under an applied magnetic field of 20 Oe along the $c$ axis. Inset: enlarged view near $T_c$. (c) Specific heat of a bulk 3R-NbSe$_2$ crystal plotted as $C_p/T$ versus $T$, measured in zero magnetic field. An equal-area construction (light blue shading) yields $T_c$ = 5.95 K and $\Delta C/(\gamma T_c)$ = 1.52, confirming bulk superconductivity. Inset: $C_p/T$ vs $T^2$ measured at $H$ = 5 T in the normal state, fitted to extract the Sommerfeld coefficient $\gamma$ and the phonon coefficient $\beta$. (d) Temperature dependence of the in-plane upper critical field for a bulk 3R-NbSe$_2$ crystal. The in-plane upper critical field exceeds the Pauli paramagnetic limit $H_P$, indicating the persistence of strong ISOC.

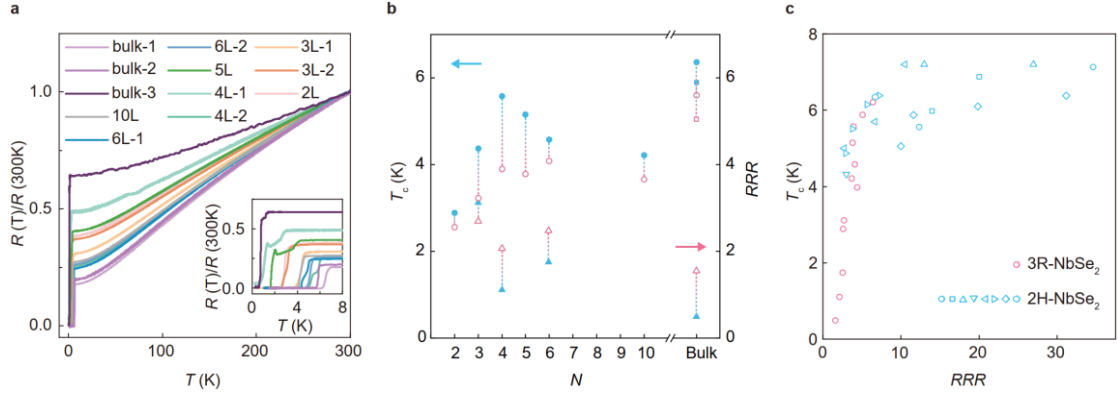

**Figure 3. Disorder-dominated superconductivity in bulk and few-layer rhombohedral NbSe$_2$.**
**(a)** Temperature-dependent resistance $R(T)$ of 3R-NbSe$_2$ samples with different thicknesses, measured from 300 K down to 0.3 K and normalized to the resistance at 300 K.. The superconducting transition temperature $T_c$ shows no systematic dependence on layer number. Inset: enlarged view of the low-temperature region. **(b)** Dependence of $T_c$ (blue symbols) and residual resistivity ratio (RRR, pink symbols) on layer number. While $T_c$ exhibits little systematic variation with thickness, RRR varies substantially among samples. Different symbols denote different devices; dashed lines highlight the correlation between $T_c$ and RRR for individual samples. **(c)** Correlation between $T_c$ and RRR for NbSe$_2$ with different crystal structures. For 3R-NbSe$_2$ (pink symbols), $T_c$ exhibits a pronounced, nearly linear dependence on RRR across samples of different thicknesses. In contrast, superconductivity in 2H-NbSe$_2$ is comparatively insensitive to RRR (blue symbols, data from references[14,16-22]).

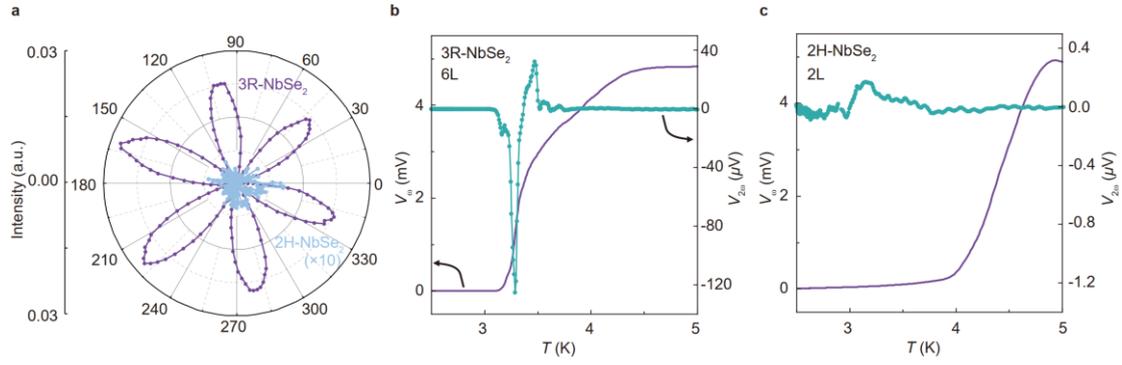

**Figure 4. Nonlinear optical and electrical responses in rhombohedral NbSe$_2$.** **(a)** Optical second-harmonic generation (SHG) signals measured at room temperature for bulk 3R-NbSe$_2$ (purple) and 2H-NbSe$_2$ (blue). The strongly enhanced SHG signal with six-fold symmetry in 3R-NbSe$_2$ reflects the absence of global inversion symmetry in the rhombohedral stacking. In contrast, the SHG signal of bulk 2H-NbSe$_2$ remains weak, consistent with the restoration of inversion symmetry. **(b)** Nonlinear electrical response in the superconducting state of a six-layer 3R-NbSe$_2$ device. The second-harmonic voltage $V_{2\omega}$ (green) appears within the superconducting transition window (purple) and rapidly diminishes both above and below $T_c$, indicating that the nonlinear response is intrinsically associated with superconductivity. **(c)** Second-harmonic voltage $V_{2\omega}$ measured in a bilayer 2H-NbSe$_2$ device under comparable conditions. The signal is more than two orders of magnitude smaller than that observed in 3R-NbSe$_2$, highlighting the role of stacking-induced inversion-symmetry breaking.

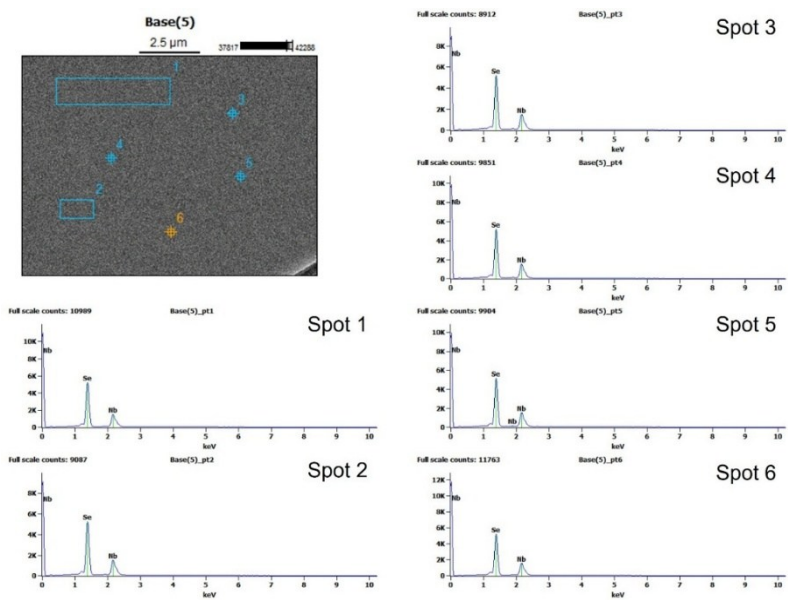

Figure S1. Results of the energy dispersive X-ray spectrum (EDS) measurements measured over multiple spots on the 3R-NbSe$_2$ crystal.

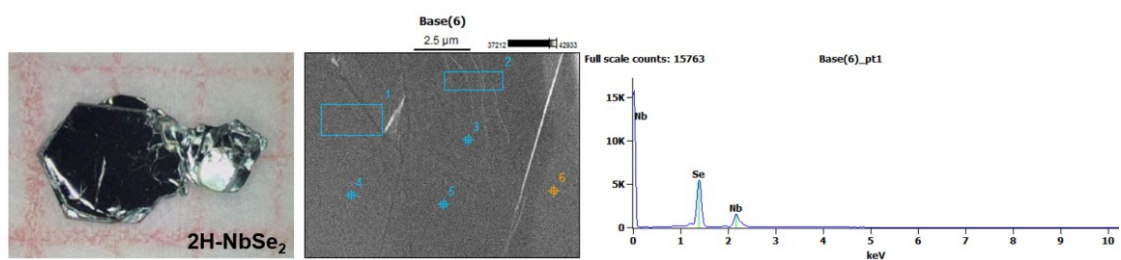

Figure S2. Results of the energy dispersive X-ray spectrum (EDS) measurements on different points of the 2H-NbSe$_2$ crystal.

Table S1. The average values of the EDS results of the 3R-NbSe$_2$ single crystal, indicating near-stoichiometric compositions as Nb : Se = 1.128 : 2.

| Element Number | Element Name | Atomic Conc. | Weight Conc. |
|---|---|---|---|
| 41 | Niobium | 36.06 | 39.89 |
| 34 | Selenium | 63.94 | 60.11 |

Table S2. The average stoichiometric composition of Nb and Se from EDS results among different 3R-NbSe$_2$ and 2H-NbSe$_2$ single crystals.

| | 3R-NbSe$_2$ | 2H-NbSe$_2$ |
|---|---|---|
| | Nb$_{1.12}$Se$_2$ | Nb$_{1.042}$Se$_2$ |
| | Nb$_{1.126}$Se$_2$ | Nb$_{1.042}$Se$_2$ |
| | Nb$_{1.14}$Se$_2$ | |
| **Average stoichiometric composition** | Nb$_{1.128}$Se$_2$ | Nb$_{1.042}$Se$_2$ |

Table S3. Crystal information and structure refinement results for NbSe$_2$ at 293 K.

| | |
|---|---|
| Empirical formula | NbSe$_2$ |
| Formula weight | 250.85 |
| Temperature | 293 K |
| Crystal system | Rhombohedral |
| Space group | *R3m* |
| Unit-cell dimensions | $a$=3.472(3) Å, $b$=3.472(3) Å, $c$=18.86(3) Å |
| | $\alpha$=90°, $\beta$=90°, $\gamma$=120° |
| Volume Z | 196.9(5) Å$^3$ |
| F(000) | 327.0 |
| Radiation | Mo K$\alpha$ ($\lambda$=0.71073 Å) |